# Learning from mistakes: The effect of students' written self-diagnoses on subsequent problem solving

Andrew Mason[1], Edit Yerushalmi[2], Elisheva Cohen[2], and Chandralekha Singh[3]

University of Central Arkansas[1], Weizmann Institute of Science[2], and University of Pittsburgh[3]

## Introduction

Helping students learn to think like a physicist is an important goal of many introductory physics courses. One characteristic of physics experts is that they have learned how to learn and use problem solving as an opportunity for learning.[1-6] In particular, physics experts automatically reflect upon their mistakes in their problem solution in order to repair, extend and organize their knowledge structure.[3-6] Unfortunately, for many students, problem solving is a missed learning opportunity.[3-6] Without guidance, students often do not reflect upon the problem solving process after solving problems in order to learn from them, nor do they make an effort to learn from their mistakes after the graded problems are returned to them. However, instruction can explicitly prompt students to diagnose their mistakes by rewarding them for such activities, so that they learn to make use of problem solving as a learning opportunity.[3-6]

Research on worked-out examples suggests that students who self-explain or elaborate to themselves what they are learning from those examples have significantly higher learning gains on tests, through mechanisms of generation of inferences and self-repair.[1] Self-repair refers to acknowledging and attempting to resolve conflicts between one's mental model and scientific models. Furthermore, self-explanation can be enhanced through explicit training in self-explaining, which is most advantageous for students with low prior knowledge.[2] It is reasonable to assume that self-repair and inference generation processes may lead to learning not only when students learn from worked examples provided to them, but also when reviewing their own solutions.

Self-diagnosis activities aim at fostering learning by providing students with time to diagnose their mistakes and rewarding them for presenting written diagnoses of their solutions to quiz problems (i.e., asking them to identify where their mistakes are, and explain the nature of those mistakes)[3-4]. Students may gain a new perspective on their own solutions by asking themselves reflective questions, in order to try to clarify what they did incorrectly and contemplate how to solve the problem correctly making use of the resources available to them.

We investigated how well introductory students self-diagnose their mistakes in physics problems when provided with time and reward for written self-diagnoses of their solutions to quiz problems, as well as the effects of their self-diagnosis on subsequent problem solving on similar problems. This approach is similar in spirit to the approach of quiz corrections[5].

# Methodology

Our study focuses upon an algebra-based introductory physics class with about 200 college students at a typical state university in the US. Recitation sections were divided into a control group and three intervention groups in which different levels of guidance were provided for self-diagnosis activities.[3,4] A majority of the students in this course were bio-science majors and pre-medical students for whom introductory algebra-based physics is mandatory. All students were taught by the same instructor in the traditional format in a large lecture hall for three hours per week. There was an hour of recitation each week and the two hundred students were divided into recitation classes with less than 50 students each. This study was implemented in the recitation sections with the consent of the course instructor and with the help of two teaching assistants.

In the control group, the instructor discussed a problem that was on a quiz the previous week, but did not ask students to diagnose their mistakes. In the three different intervention groups, or self-diagnosis groups, students were asked to diagnose mistakes on photocopies of their previous week's quiz solutions to the problem, i.e., they were asked to identify where they went wrong and explain the nature of their mistakes. The self-diagnosis groups differed in the aids students were given for the self-diagnosis activities. In group I (for "Instructor"), the instructor presented an outline of the solution orally, skipping details of the derivations. Students filled in a self-diagnosis rubric focused on deficiencies in their physics knowledge as well as in their approach to solving the problem in a systematic manner. An example of a student's written solution and self-diagnosis from group I can be seen in Figure 1. In group W (for "Written solution"), students were provided with a detailed written solution including the reasoning for various solution steps. In group S (for "Struggle"), students were only given the final answer to the problem, and they were allowed to use their notes and textbooks for self-diagnosis, but otherwise had to struggle on their own to determine their errors (the problem could not be solved using an algorithmic approach by most students). We note that there were five recitation sections for the course: two were randomly assigned to be the control group, and the other three to be group I, group W or group S. Upon completion of the self-diagnosis activity, all students were provided a detailed solution for the problem they self-diagnosed. A "transfer" problem, with the same underlying physics principles as the quiz problem students self-diagnosed, was then posed in a midterm exam. The goal of administering the transfer problem was to investigate the extent to which students in different intervention groups differed from the control group, in solving problems with similar underlying physics principles, when there was no support provided to any of the groups.

Table 1 provides a summary of the different groups and the respective levels of guidance given to each group. We note that in the control group, the instructor discussed in great detail the solution of the quiz problem with the students when the graded quizzes were returned in the next recitation class. Other groups did not get to hear the instructor's detailed explanation of the correct solution for the quiz problem (except intervention group I in which a brief outline was provided before students diagnosed their mistakes). Thus, the time on task was essentially the same in the control group as in the intervention groups, but the task was more passive: listening to the instructor's detailed explanation of the correct solution rather than diagnosing mistakes.

Figure 1: Written solution and self-diagnosis from a sample student from group I.

|  | Control group | Intervention groups | | |
| --- | --- | --- | --- | --- |
|  |  | I | W | S |
| Initial training | The training follows a modified version of the intervention sequence: Students are given, for demonstration purposes, an incorrect solution of a "training problem" and diagnose it according to the procedure specified by their intervention group. Then the TA demonstrates how the incorrect solution should be diagnosed for the training problem. | | | |
| Quiz | Students solve the quiz problem | | | |
| Intervention | TA discusses solution of the quiz problem | TA presents in class an outline of the solution. Students self-diagnose their solutions: circle mistaken parts, explain, and sort mistakes in a self-diagnosis rubric | TA provides a written worked-out example. Students self-diagnose their solutions: they circle mistaken parts and explain the mistakes | Students can use their notes and textbooks. Students write their self-diagnosis |
| Midterm Exam | Students are given a transfer problem, paired to the quiz problem, on midterm examination | | | |

Table 1: Summary of different interventions for each group.

We note that before students in all intervention groups were asked to diagnose their own mistakes, they were given training in how to diagnose mistakes; they were provided with a hypothetical student solution, in which common mistakes were planted strategically, and asked to diagnose the mistakes in that solution. After students had diagnosed the mistakes in that solution to the best of their ability, they were given an example of a good diagnosis of mistakes for that solution, followed by a discussion about why the example was a good way to diagnose mistakes.

To examine the effect of prior knowledge on student learning from self-diagnosis, we carried out the experiment twice: first in the context of a challenging quiz problem, and then in a less challenging, more typical quiz problem. Both problems are multi-part problems. In the challenging problem, a person comes down from a certain height on a roller coaster and then goes over a bump, which is shaped like an arc of a circle. Students were asked to find the normal force on the person when passing over the top of the bump. Students must find the speed of the person at the top of the bump using conservation of mechanical energy, and then apply Newton's Second Law to this non-equilibrium situation with centripetal acceleration to find the normal force on the person at the top of the bump. In the less challenging problem, students had to use both conservation of mechanical energy and conservation of momentum to find the speed of two objects that merge after a completely inelastic collision and move up a curved ramp.

## Results

We looked for correlation between the quiz score and the transfer problem score for each group (with or without self-diagnosis). We also examined the effect of self-diagnosis activities on students' ability to self-repair their understanding (see reference 4). In the different intervention groups, we investigated how well students self-diagnosed mistakes in their solutions of quiz problems, and examined the correlation between students' self-diagnosis performance and subsequent problem solving on a transfer midterm exam problem.

**Correlation between the initial quiz score and score on the transfer problem:** For both challenging and typical problem settings, we found significant correlations between students' score on the quiz problem and on the transfer problem in the control group (correlation = 0.41, p-value = 0.03), suggesting that the gap between students with low and high prior knowledge remained unchanged. In contrast, we found no significant correlations between the quiz score and score on the transfer problem in any of the self-diagnosis groups in either setting, suggesting, as further evidence has confirmed, that all of the self-diagnosis interventions reduced the gap between students with low and high prior knowledge to some extent. We conclude that the self-diagnosis activity enhanced self-repair in students with low prior knowledge.

**Correlation between self-diagnosis score on the typical problem and score on subsequent transfer problem**: In the context of the typical problem situation, we find that the self-diagnosis score was correlated (correlation = 0.54, p-value = 0.01) with subsequent problem-solving performance on the transfer problem, but only for group S, when textbooks and notes were the sole means of guidance available to the students to help them with self-diagnosis. We hypothesize that group S students experienced more cognitive involvement in their self-diagnostic activity than group I or group W students did. Group S students had to struggle to find and identify information in their notes or textbooks that was relevant to the problem to be diagnosed in order to self-diagnose their mistakes. However, in the typical problem setting, group S students were able to self-diagnose their mistakes. As a result, students who provided an "acceptable" self-diagnosis and obtained good self-diagnosis scores were often those who engaged in self-repair. The fact that in other interventions, there was no correlation between self-diagnosis score and score on subsequent transfer problem may be because many students in those interventions may have done superficial self-diagnosis and they may not have cognitively engaged in the self-diagnosis of their mistakes. In particular, some students in those interventions may have simply compared the surface features of their solution with the solution provided for self-diagnosis, thus providing an "acceptable" diagnosis and obtaining a good score on the self-diagnosis task without actually engaging in self-repair. Accordingly, the self-diagnosis score would not correlate with performance on transfer problems for those interventions.

**Performance on the transfer problem of group S on the challenging problem:** In the context of the challenging problem, we find that there is no correlation between the self-diagnosis score and the subsequent transfer problem score for any group; however, group S students, who struggled to diagnose their mistakes with only textbooks and notes in class, obtained a group average of 46% higher scores on the subsequent transfer problem than did group W students, who were asked to diagnose their mistakes using a detailed solution of the challenging problem provided to them (averages for these two groups, including all students in the group, are 51% for group S vs. 35% for group W). One possible reason is that students who struggled the most during self-diagnosis, i.e. group S, were more motivated to learn how to solve that type of problem correctly when the entire class was provided the correct solution after the self-diagnosis activity was over, but before the subsequent transfer problem was posed in the midterm exam. Thus, struggling during self-diagnosis may have had a positive effect on what students did with the detailed solution provided to all students after the self-diagnosis.

## Summary and Conclusions

Self-diagnosis activities ask students to diagnose their mistakes after having solved a problem on their own. Self-diagnosis tasks are aimed at fostering expert-like diagnostic behavior of learning by explicitly requiring students to diagnose their mistakes in their solutions. We investigated students' ability to self-repair their understanding when reflecting on their own solutions to a challenging and a typical problem, using common resources such as providing students with the instructor's outline, detailed solutions, or letting them use their textbooks and notes. Student self-diagnoses can be used as a means for formative assessment,[7] and can provide useful feedback to both the instructor and the students about what students are able to do at a given time in the course, so that appropriate improvements can be made. This method of formative assessment is compatible with other methods instructors may be using.

Our study suggests that students with low prior knowledge benefit the most from self-diagnosis activities, and the gap between high and low performers on the initial quiz shrinks on the transfer problem after the self-diagnosis in all groups; however, the gap remains if students do not perform self-diagnosis. Those with high scores remain fairly high, as they are self-selected as good learners, and they saturate the high end of the scale since there is not a lot of space to improve. In the context of the typical problem situation, we find that the self-diagnosis score was correlated with subsequent performance on the transfer problem only for group S, for whom textbooks and notes were the sole means of guidance available to the students to help them with self-diagnosis. Group S students struggled the most but were still able to do an acceptable self-diagnosis. Our research also suggests that struggling may be beneficial for successful future learning, and students may be more motivated to engage with instructional material in a more meaningful way after the struggle. For example, for the challenging problem, group S students were unable to do adequate self-diagnosis in class with textbook and notes, but performed 46% higher on the transfer problem than did students who were given the detailed correct solution during self-diagnosis task. One possible reason for this difference may be that group S students, who struggled to self-diagnose, were more cognitively engaged with the detailed solution provided after they had struggled with the self-diagnosis process with textbooks and notes than group W students, who were provided detailed solution during the self-diagnosis task.[8-9] It is important for instructors to help students understand the importance of cognitive engagement and the role of appropriate struggle in learning physics. The study highlights that an important challenge for instructors is finding the right balance between limiting support to allow deep-level engagement and providing support to allow students to connect what they are learning with their prior knowledge.